\documentstyle[prc,aps,psfig]{revtex}
\newcommand{\be}{\begin{eqnarray}}
\newcommand{\ee}{\end{eqnarray}}

\newcommand{\bi}{\begin{itemize}}
\newcommand{\ei}{\end{itemize}}
\begin{document}
\twocolumn[\hsize\textwidth\columnwidth\hsize
           \csname @twocolumnfalse\endcsname
\title{Transport with three-particle interaction}
\author{Klaus Morawetz}
\address{LPC-ISMRA, Bld Marechal Juin, 14050 Caen
 and  GANIL,
Bld Becquerel, 14076 Caen Cedex 5, France\\
Max-Planck-Institute for the Physics of Complex Systems, 
Noethnitzer Str. 38, 01187 Dresden, Germany}
\maketitle
\begin{abstract}
Starting from a point - like two - and three - particle interaction the
kinetic equation is derived. While
the drift term of the kinetic equation turns out to be determined by the known
Skyrme mean field the collision integral appears in two -
and three - particle parts. The cross section results from the same
microscopic footing and is naturally density dependent due to the
three - particle force. By this way no hybrid model for drift and
cross section is needed for
nuclear transport. Besides the
mean field correlation energy the resulting equation of state has  also a two - and three - particle
correlation energy which are both calculated
analytically for the ground state. These energies contribute to the equation of
state and lead to an occurrence of a maximum at 3 times nuclear density
in the total energy.
\end{abstract}
\vskip2pc]

\section{Introduction}

The equation of state of nuclear matter is known to saturate at the
density of $n_0=0.16$fm$^{-3}$ with a binding energy of $-16$MeV. Refined two-particle calculations using Bruckner theory
\cite{K74} and beyond \cite{HMPO96} are not
crossing the ``Coester line'' in that they lead to binding energies
and/or densities
above the saturation ones. Only density dependent Skyrme
parameterizations originally introduced by
Skyrme \cite{S56,S59}, three-particle interactions \cite{CPW83,JRK83} or relativistic treatments \cite{BFF99} can
reproduce the correct binding energy and saturation density of nuclear matter. The Skyrme parameterizations are
derived from an effective three-particle interaction \cite{VB72}. This
leads to a nonlinear density dependence in the effective three -
particle part which is responsible for saturation. The importance of three-particle
collisions in nuclear matter transport has been demonstrated, e.g. 
in \cite{BGM94}. The density
dependence deviating from that arising by
three-body contact interaction has been introduced and compared with
experiments in \cite{Ko75}.  

The relativistic approach on the other
hand yields immediately the correct saturation with two - particle
exchange interactions. This has been traced down to the nonlinear
density dependence of scalar density and consequently nonlinear
mean fields \cite{BGM87} which leads to a density contribution to the
binding energy of $(n/n_0)^{8/3}$.
Higher order effects have been shown to lead to a $(n/n_0)^{3.4}$
dependence, see \cite{BWBS87} and citations therein. 
Therefore the physics of the saturation mechanism is certainly
a nontrivial density dependence of the mean field. Though 
the Skyrme interaction has been overwhelmingly successful and the
evidence of saturation by three-particle interaction
\cite{CPW83,JRK83}, 
it is
puzzling that a transport
theory with three-particle contact interaction has not been
formulated. In this paper we will derive the corresponding kinetic
equation using nonequilibrium
Green's function  and we will show that the three-body interaction term leads to
a $(n/n_0)^{10/3}$ contribution to the binding energy.

The transport theory including three - particle interactions
seems to be of wider interest e.g. in
\cite{SRS85}
it has been found that the three-body interactions
have a measurable influence on thermodynamic properties of fluids in equilibrium.
The description of nonequilibrium in nuclear matter is mostly based on
the 
Boltzmann (BUU) equation which
uses the Skyrme parameterization to determine the mean field and drift
side of the kinetic equation. The collisional integral is then calculated
with a cross section which arises from different theoretical
impact. In addition one usually neglects  the three-particle collision
integral. These hybrid models have worked quite successfully despite their weak rigour in microscopic foundation.

In this paper the transport equation will be derived for
two- and three - particle contact interactions. 
We will see that a natural density
dependent mean field appears, in agreement with variational
methods. Further we will obtain, on the same ground, density dependent
cross sections. This has the advantage that we can derive a BUU
equation which has the same microscopic impact on the drift and
collisional side. Finally the three -particle collision integral
appears naturally from this treatment. From balance equations we will
derive the density dependent energy which gives the equation of state.

\section{Three - particle Kadanoff and Baym equation}

We consider a Hamilton system with the Hamiltonian
\be
H&=&\sum\limits_ia_i^+{\hbar^2 \nabla^2\over 2 m} a_i \nonumber\\
&&+\frac 1 2 \sum\limits_{ij}a_i^+a_j^+a_ia_jV_2(i,j)
\nonumber\\
&&+\frac 1 6 \sum\limits_{ijk}a_i^+a_j^+a_k^+a_ia_ja_kV_3(i,j,k)
\ee
where $a_i^+, a_i$ are creation and annihilation operators with
cumulative variables $i=(x_i,t_i,...)$. We assume
the potential is contact - like as
\be
V_2(i,j)&=&t_0\delta(i-j)\nonumber\\
V_3(i,j,k)&=&t_3\delta(i-j)\delta(i-k).
\label{2}
\ee
The Heisenberg equations of motion for the annihilation operators read
\be
i\hbar\partial_{t_1} a_1&=&[a_1,H]\nonumber\\
&=&-{\hbar^2 \nabla_1^2\over 2 m} a_1+\sum\limits_2 V(1,2)a_2^+a_1a_2
\nonumber\\
&&+\frac 1 2 \sum\limits_{23}V(1,2,3)a_2^+a_3^+a_1a_2a_3.
\label{mot}
\ee
From this we get the equation of motion for the causal Green function
$G(1,2)=\frac 1 i <T a_1^+a_2>$ with  the time ordering $T$ and the
average taken  about the nonequilibrium density operator
\be
&&(i\hbar \partial_{t_1}-{\hbar^2 \nabla_1^2\over 2 m})G(1,1')=\delta(1-1')\nonumber\\
&&+\int d2 V(1,2) G_2(1,2,1',2^+)
\nonumber\\&&
+\frac 1 2 \int d2 d3 V(1,2,3) G_3(1,2,3,1',2^+,3^+)\nonumber\\
&&\equiv \delta(1-1')-\int dx_2\Sigma_{HF}(x_1,x_2,t)G(x_2,t,x_1,t)
\nonumber\\&&
+\int d2 (\Sigma(1,2)G(2,1')-\Sigma^>(1,2)G^<(2,1'))
\label{MS}
\ee
where the $1^+$ denotes the space - time point $1=(x_1,t_1+\epsilon)$
with a time infinitesimally larger than $1$.
This equation is enclosed in the standard manner introducing the self
energy $\Sigma$.
We have split the mean field parts $\Sigma_{HF}$ including exchange
and have introduced for the rest the self energy on the Keldysh contour. This is equivalent to the weakening of initial correlations. The correlation functions are
$G^<(1,2)=<a_2^+a_1>$ and $G^>(1,2)=<a_1a_2^+>$ respectively and their
equation of motion follows from (\ref{MS}) in the form of the known
Kadanoff and Baym equation \cite{KB62,LSV86}
\be
i(G_0^{-1} G^<-G^< G_0^{-1})=\Sigma^RG^<-G^<\Sigma^A
\label{KB}
\ee
where one takes advantage of the retarded and advanced functions
$G^{R/A}=\mp i\theta[\pm(t_1-t_2)](G^>\mp G^<)$ and understands
products as integration over inner variables. 

\begin{figure}
\psfig{file=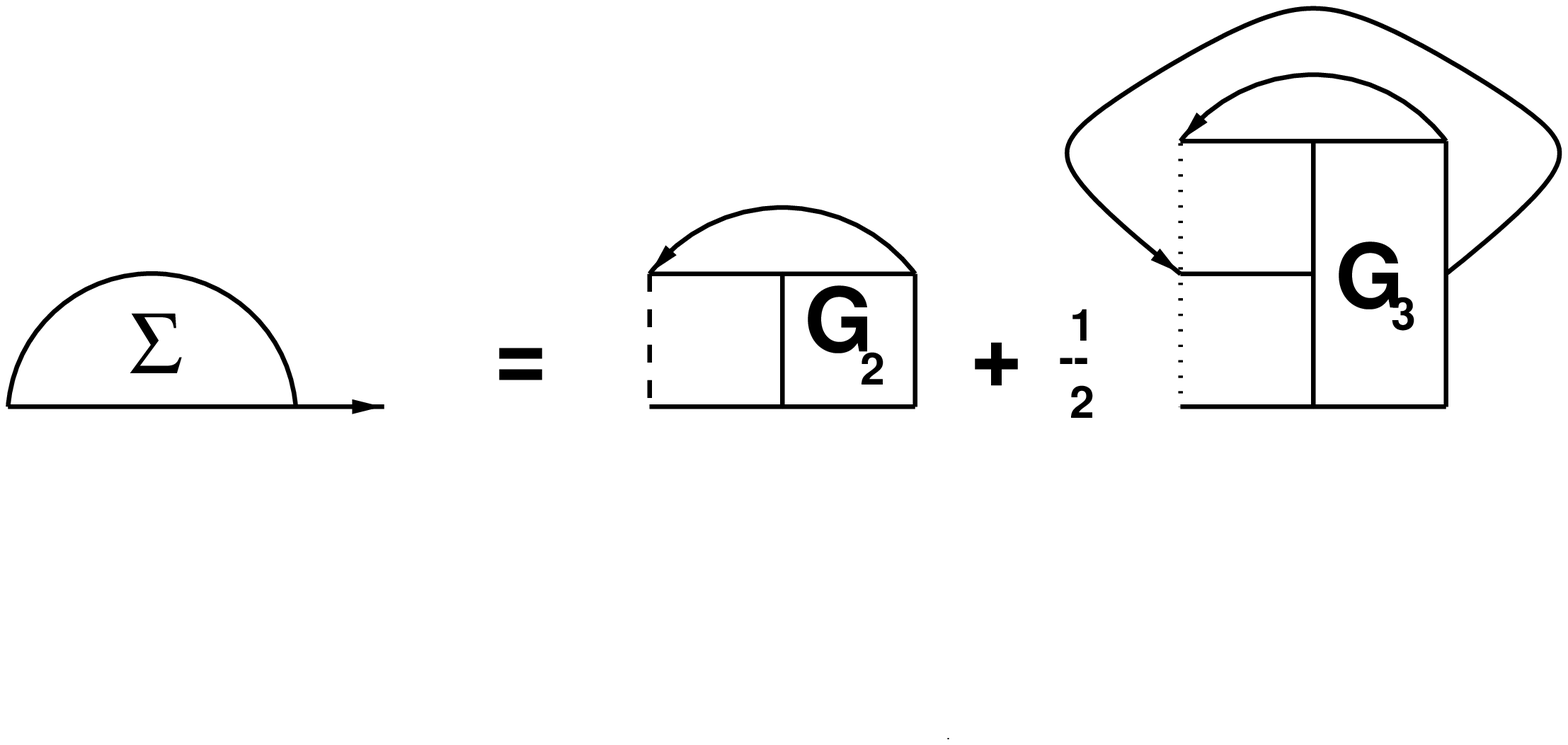,width=8cm}
\caption{The definition of self energy. We denote the two-particle interaction by long dashed lines and the three-particle interaction by short dashed line.}\label{f1}
\end{figure}

\subsection{Mean field parts}
\begin{figure}
\psfig{file=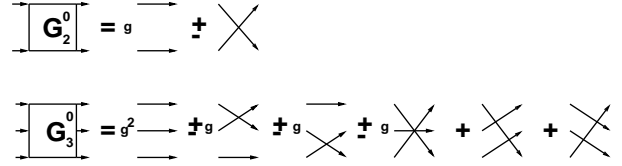,width=8cm}
\vspace{2ex}
\caption{The lowest order expansion of Green functions. The
  statistical factor $g$ due to degeneracy in the system is indicated 
explicitly.}\label{f2}
\end{figure}

We first calculate the mean field parts. These parts come from the first order interaction diagrams of figure \ref{f1}. To this end we use the two- and three- particle Green function in lowest order approximation seen in figure \ref{f2}.

Introducing these diagrams into figure \ref{f1}, one obtains for contact interaction (\ref{2})
\be
&&\Sigma^{\rm HF}(1,1')=\Biggl [{g\pm 1\over g} t_0 n(1)+{(g\pm 1)( g\pm2)\over 2 g^2}t_3 n(1)^2 \Biggr ] \delta(1-1')\nonumber\\&&
\label{hf}
\ee
where we have used the fact that the density is
$n(1)=G^<(1,1)/g$. Here and in the following we write the upper sign
for Bosons and the lower sign for Fermions. The
result (\ref{hf}) is the known Skyrme mean field expression in nuclear
matter for $g=4$. It resembles an effective density dependent
two-particle interaction arising from three-particle contact
interaction. As a consistency check we see that for $g=2$ degenerated
Fermionic system, like spin -1/2 Fermions, no three-body term appears.
Pauli-blocking forbids  two particles to meet at the same
point with equal quantum numbers which one would have for three -
particle contact interaction and degeneracy of $g=2$.

\subsection{Kinetic equation}

\begin{figure}
\psfig{file=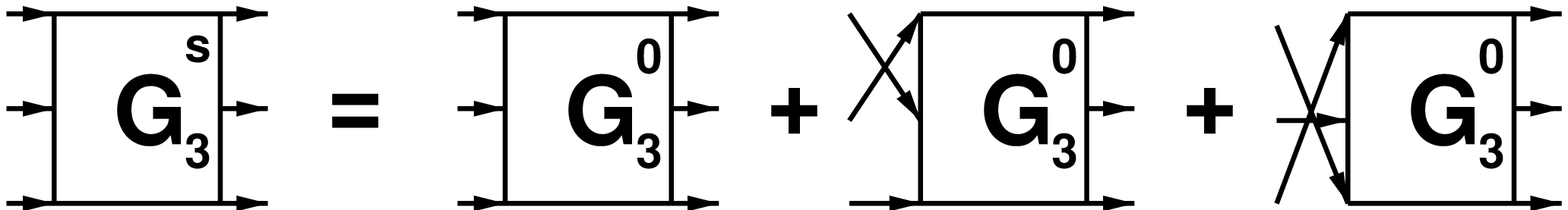,width=8cm}
\vspace{2ex}
\caption{The symmetrized three-body Green function with respect to 
incoming lines which are used in figure \protect\ref{g31}.}
\label{g3anti}
\end{figure}

For the kinetic equation we need as the lowest order approximation
the next diagrams of figure \ref{f2} including one interaction line. 
While there are different expansion techniques at hand
\cite{D90,SRS85} we will perform here a scheme as close as possible to
the causal Green's function including all exchanges
which are presented in figure \ref{f2}. This will give us the
advantage that we can consider all diagrams which differ by exchange
of outgoing lines as equal. To this aim we introduce the abbreviated symmetrized
Green's function with respect to incoming lines in figure
\ref{g3anti}.

\begin{figure}
\psfig{file=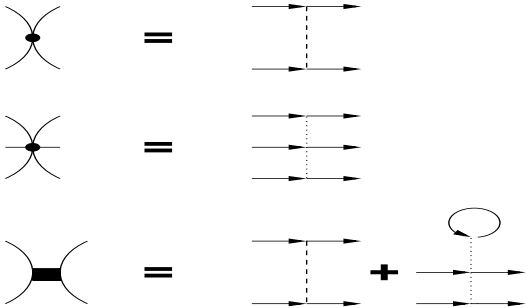,width=7cm,height=4cm}
\vspace{2ex}
\caption{The introduced vertices. The three - and two - particle
  interaction potentials are the same as in figure\protect\ref{f1}.}\label{gvert}
\end{figure}

\begin{figure}
\psfig{file=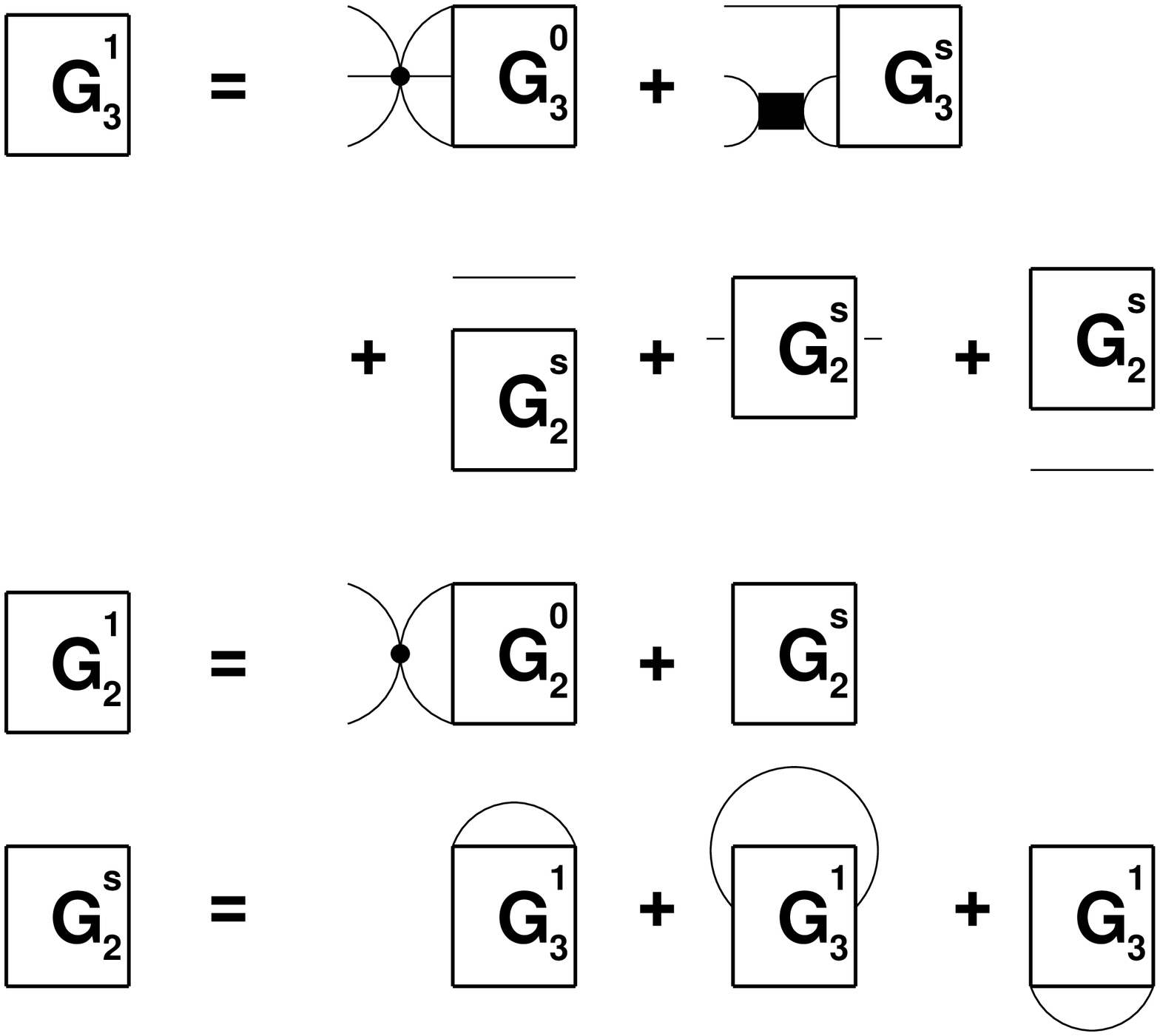,width=8cm}
\vspace{2ex}
\caption{The first order expansion of the two - and three - particle Green's functions with respect to interaction. }\label{g31}
\end{figure}

The expansion of the three -body Green's function is given in figure
\ref{g31}. Besides the three -particle interaction potential we have
to consider all possible single line interactions between two incoming
lines. This is shortened by the introduction of the symmetrized Green
function in figure \ref{g3anti}
and the vertices of figure \ref{gvert}.

\begin{figure}
\psfig{file=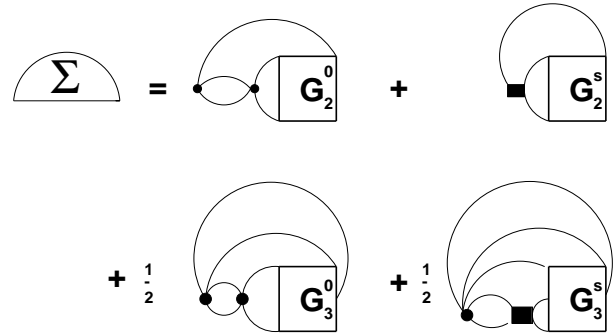,width=8cm}
\vspace{2ex}
\caption{The self energy according to figure\protect\ref{f1} with the
  expansion of figure \protect\ref{g31}.}\label{gself}
\end{figure}

Since we can also have nontrivial combinations of two incoming lines
and one free line for the three - body Green function
we have introduced an auxiliary two - body Green function $G_2^{\rm s}$ which is
defined in figure \ref{g31}. Please note that the exchange of
outgoing lines does not lead to distinguished diagrams. The two -
particle Green function is then given by figure \ref{g31}.
Introducing now this expansion into the definition of self energy
in figure \ref{f1} we obtain the 4 diagrams of figure \ref{gself}.
All other combinations lead either to disconnected diagrams or to
equivalent ones by interchanging outgoing lines. Here the seemingly
disconnected diagrams for $G_3^1$ in figure \ref{g31} lead together
with the corresponding diagrams from $G_2^1$ to the vertex in front of
$G^s_2$ of figure \ref{gself}.

\begin{figure}
\psfig{file=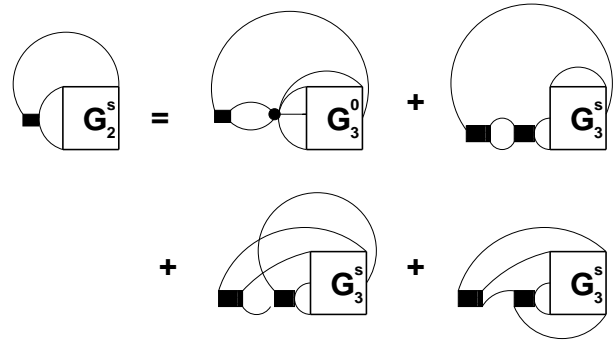,width=8cm}
\vspace{2ex}
\caption{The enclosing diagram of the auxiliary Green function $G_2^s$.}\label{g2anti}
\end{figure}

The enclosing diagram about the auxiliary Green function  $G_2^s$ is
given in figure \ref{g2anti}. Please denote that here the seemingly
disconnected diagrams of figure \ref{g31} vanish. We have used again
the fact that the interchange of outgoing lines does not lead 
to distinguished diagrams.

Inserting figure \ref{g2anti} in figure \ref{gself} we obtain the
final result for the self energy in the  Born approximation
\be
&&\Sigma^<(1,1')=(g\pm1) \left [t_0^2\right .\nonumber\\
&&+\left . (g\pm2)\left ( 
4 t_0^2 +
6 t_3^2 {n(1)n(1')
\over g^2}+5t_0t_3 ({n(1)\over g}+{n(1')\over g}) \right )
\right ] 
\nonumber\\&&\times G^<(1,1')^2G^>(1'1)\nonumber\\
&&+ \frac 1 2 (g\pm 1)(g\pm 2) t_3^2 \,\,G^<(1,1')^3G^>(1'1)^2.
\label{born}
\ee
Note that all interactions vanish for the fermionic case
$g=1$ since the Pauli-principle does not allow contact interaction in
this case. For $g=2$ only the two - particle interactions survive for
fermions as one expects from the Pauli principle. That the result is
symmetric in the densities at the two time-space points needs no further comment.

From the self energy we can write down the Kadanoff and Baym equation (\ref{KB})
in closed form. If we employ standard gradient expansion and convert
the equation (\ref{KB}) into an equation for the pole part of the Green's function we obtain the final kinetic equation
\be
\label{kin}
&& \frac{\partial}{\partial t} f_1(t) +{p_1\over m}{\partial \over
\partial r} f_1(t)-{\partial\over \partial r}\sigma^{\rm
HF}(t){\partial\over \partial p_1}f_1(t)
     \nonumber\\
&=& \frac{2}{\hbar ^2}
     \int \frac{dp_2 dp_1' dp_2'}{(2 \pi \hbar )^6}
     T_2^2
     \delta (p_1 + p_2 - p_1' - p_2')
     \nonumber \\
  &\times& \int_{t_0}^{t }d\tau {\rm cos}
\left [\frac{1}{\hbar} (E_1 + E_2 - E_1' - E_2')(t-\tau)
\right ]\nonumber\\
  &\times &
\left \{f_1'( \tau ) f_2'(\tau )
{\bar f_1}( \tau )
{\bar f_2}( \tau )- f_1( \tau ) f_2( \tau ) {\bar f_1'}(
\tau )
{\bar f_2'}( \tau ) \right \}
     \nonumber \\
&+& \frac{2}{\hbar ^2}
     \int \frac{dp_2 dp_3 dp_1' dp_2'dp_3'}{(2 \pi \hbar )^{12}}
     T_3^2
     \delta (p_1\! +\! p_2\! +\!p_3\!-\! p_1' \!-\! p_2'\!-\!p_3')
     \nonumber \\
  &\times& \int_{t_0}^{t }d\tau {\rm cos}
\left [(E_1 + E_2 +E_3- E_1' - E_2'-E_3'){t-\tau\over \hbar}
\right ]\nonumber\\
  &&\times 
\left \{ f_1'( \tau ) f_2'(\tau )       f_3'( \tau )
{\bar f_1}( \tau )
{\bar f_2}( \tau ){\bar f_3}( \tau )\right .\nonumber\\
&-& \left .f_1( \tau ) f_2( \tau )f_3( \tau ) {\bar f_1'}(
\tau )
{\bar f_2'}( \tau ){\bar f_3'}( \tau )\right \}
     \nonumber \\
\end{eqnarray}
with $\bar f = 1-f$, and the particle dispersion
$E=p^2/2m+\sigma^{\rm HF}$.
The quasiparticle distribution functions are normalized to the
density as $g \int {d p\over (2 \pi \hbar)^3} f(p)=n$. For the sake of
simplicity we have suppressed the notation of centre of mass space
dependence. Neglecting the retardation in the
distribution functions and taking $t_0\rightarrow \infty$ gives the standard
Boltzmann two- and three- particle collision integrals.

The introduced two - and three- particle T-matrices are read off from
(\ref{born}) as first Born approximation
\be
T_2^2
&=&{g\pm1\over g^2} \left [t_0^2\right .\nonumber\\
&+&\left . (g\pm2)\left ( 
4 t_0^2 +
6 t_3^2 {n(1)n(1')
\over g^2}+5t_0t_3 ({n(1)\over g}+{n(1')\over g}) \right )\right]
\nonumber\\
T_3^2&=&\frac 1 2 {g\pm 1\over g^3}(g\pm 2) t_3^2.
\nonumber\\
\ee
The reader is remind that we have generally to sum over collision partners and to
average over the outgoing products. While the summation is already
taken into account explicitly in the diagrammatic approach above 
we had to divide by $g^2$ and $g^3$ for
the two- and three - particle outgoing collision partners
respectively. An effective cross section can be defined in the usual
way
and read up to constants
\be
{d \sigma_2 \over d \Omega}&\sim& T^2_2= {3\over 16} \left [9
  {t_0^2}+\frac 3 4 t_3^2 n^2+ 5 t_0 t_3 n \right ]\nonumber\\
&&={27\over 16} \left [
  (t_0+\frac 1 6 t_3 n)^2+ \frac 2 9 t_3 n (t_0+\frac 1 4 t_3 n) \right ]
\nonumber\\
{d \sigma_3 \over d \Omega}&\sim&T_3^2= {3\over 64} t_3^2.
\label{cross}
\ee
If one compares this with the effective two- particle cross section derived
in \cite{AYG98} one sees that it
differs by the last two terms of (\ref{cross}) proportional to $t_3$. 
This is due to the fact that the underlying three - particle
process is taken into account explicitly here while in \cite{AYG98}
an effective two - particle kinetic theory has been
developed. Obviously we have to face a partial cancellation between
three - and two -
particle collisions. This will become explicit in the discussion of
correlation energies. We will see that the two - and three - particle
correlation energies indeed cancel partially concerning the $t_3$ terms.

The ratio of the effective cross sections are 
for the case of nuclear matter ($g=4$)
\be
{d \sigma_2 \over d\sigma_3}= {T_2^2\over T_3^2}
&=&4 \left [9 {t_0^2\over t_3^2}+\frac 3 4 n^2+ 5 {t_0 \over t_3} n \right ].
\label{crossr}
\ee
This could serve as the measure of relative importance of the
corresponding collision processes, but we prefer to discuss this later
in 
terms of dispersed energy by the two- and three- particle collision
integrals
since this includes also the Pauli blocking. 

We would like to point out that we have fulfilled our task and have derived
a kinetic equation where the drift as well as the collision integral
follows from the same microscopic impact. The drift represents a
Skyrme mean field and the collision side shows a two - and three - particle collision integral where the cross sections are calculated from the same parameters as the mean field.

\section{Balance equation}

By multiplying the kinetic equation with $1,p$ and $E$, one obtains the
balance for density $n$, momentum $u$ and energy ${\cal E}$. 
Since the collision
integrals vanish for density and momentum balance we get the usual
balance equations
\be
&&{\partial n\over \partial t}+{\partial \over \partial R} \int
{dp\over (2 \pi \hbar)^3} {\partial E\over \partial p}  f=0\nonumber\\
&&{\partial u_i\over \partial t} +{\partial \over \partial R_j} \int
{dp\over (2 \pi \hbar)^3} (p_i {\partial E\over \partial p_j} f+{\cal
    E}\delta_{ij})=0
\ee
where the mean field energy of the system varies as 
\be
\delta {\cal E}&=&\int
{dp\over (2 \pi \hbar)^3} {\delta {\cal E}\over \delta
    f(p)} \delta f(p)
\nonumber\\
&=&\int
{dp\over (2 \pi \hbar)^3} E \delta f(p)
\ee
such that from (\ref{hf}) follows
\be
{\cal E}=<{k^2\over 2 m}>+{g\pm 1\over 2 g} t_0 n^2+{(g\pm 1)( g\pm2)\over 6 g^2}t_3
n^3.
\label{hf1}
\ee
With the help of this quantity the balance of energy density reads
\be
&&{\partial {\cal E}\over \partial t} +{\partial \over \partial R} \int
{dp\over (2 \pi \hbar)^3} E {\partial E\over \partial p}   f=-{\partial \over \partial t}E_{\rm
    corr_2}(t)-{\partial \over \partial t}E_{\rm corr_3}(t)
\label{bal}
\nonumber\\
&&
\ee
with the two-particle correlation energy \cite{M94,MK97}
\be
&&E_{\rm corr_2}(t)\nonumber\\
&&=-\frac{g}{\hbar }
     \int\limits_{t_0}^td\tau\int \frac{dp_1 dp_2 dp_1' dp_2'}{(2 \pi \hbar )^9}
     T_2^2
     \delta (p_1 + p_2 - p_1' - p_2')
     \nonumber \\
  &&\times \sin{\left [(E_1 + E_2 - E_1' - E_2'){t-\tau \over
          \hbar}\right ] } 
%  \nonumber\\
%  &\times &
f_1'( \tau ) f_2'( \tau )
{\bar f_1}( \tau )
{\bar f_2}(\tau )
     \nonumber \\
     \label{e2corr}
\ee
and the complete analogous expression for the three particle
energy.

Please note that in the balance equations for the density and momentum
no correlated density or correlated flux appears. This is due to the
Born approximation. For more nontrivial approximations as e.g. the
ladder summation these correlated observables appear \cite{SLM96,LSM97}.

\subsection{Fit to nuclear matter ground state}

As one can see, it is not enough to use the mean field parameterization to fit the
equation of state as done in most approaches so far. Since the collision
integral induces a two - particle and three - particle correlation energy we have to take this into
account and have to  refit the parameter $t_0,t_3$ to the binding properties of nuclear
matter. To illustrate this fact let us evaluate the fit without two -
and three - particle
correlation energy and therefore without collision integrals.

\subsubsection{Hartree-Fock parameterization}
Taking into account only Hartree-Fock mean field correlations
(\ref{hf1}) the ground state energy for nuclear matter reads
\be
{E\over A}={{\cal E}\over n}=\frac 3 5 \epsilon_f +\frac 3 8 t_0
n+\frac{1}{16}t_3 n^2.
\label{hf2}
\ee
Then the nuclear
binding at $n_0=0.16$ fm$^{-3}$ with an energy of $E_B=-15.68$MeV is reproduced
by the set
\be
t_0&=&-{16\over 15 n_0}(2 \epsilon_f-5 E_B)=-1026.67 {\rm MeV\, fm^3}\nonumber\\
t_3&=&{16 \over 5 n_0^2} (\epsilon_f-5 E_B)=14625 {\rm Mev\, fm^6}\nonumber
\ee
which leads to a compressibility of
\be
K=9 n^2 {\partial^2 E/A\over \partial n^2}=\frac 9 8 t_3
n_0^2-\frac 6 5 \epsilon_f=377 {\rm MeV}.
\label{k}
\ee

\subsubsection{Parameterization with two - and three - particle
correlation energy}

Now we consider the two - particle correlation energy. From (\ref{kin})
we obtain the total energy 
\be
{\cal E}+ E_{\rm corr_2}+E_{\rm corr_3}.
\label{hf3}
\ee

We want to calculate the long time limit which represents the
equilibrium value. From the identity $\int\limits_0^{\infty} \sin{x t} d t={{\cal P}\over x}$
a principle value integration replaces the $sin$ term in
(\ref{e2corr})
\be
&&E_{\rm corr_2}(\infty)=- g
     \int \frac{dp_1 dp_2 dp_1' dp_2'}{(2 \pi \hbar
)^9}
     T_2^2{{\cal P} \over E_1 + E_2 - E_1' - E_2'}
          \nonumber \\
  &&\times \delta (p_1 + p_2 - p_1' - p_2')
f_1' f_2'
{\bar f_1}
{\bar f_2}.
\label{eq}
\ee

For ground state Fermi distributions this expression can be
integrated analytically \cite{MK97}
and we find from (\ref{eq}) the known Galitskii result for the two -
particle ground state correlation energy \cite{PN66}
\be
{E_{\rm corr_2}\over n}&=&4  \epsilon_f {2 \log
2-11\over35} ({p_f m T_2\over 4 \pi^2 \hbar^3})^2\nonumber\\
&=&-{5.79 10^{-5}\over {\rm MeV\, fm}^2} n^{4/3} \left [ 9 t_0^2+\frac 3 4 t_3^3 n^2+ 5 t_3 t_0 n \right ].
\label{e2}
\label{contact}
\ee
As pointed out in \cite{LL79} we had to subtract here an infinite value, i.e. the term proportional to $f_1 f_2$ in
(\ref{eq}). This can be understood as renormalization of the contact potential and is formally hidden in
$E_{\rm corr}(0)$ when time integrating equation (\ref{bal}). For finite range potentials we have an intrinsic cut-off due
to range of interaction and such problems do not occur.

The correlational two - particle energy (\ref{contact}) is
always negative for fermionic degeneracies $2<g<8$ and for bosonic
degeneracy $g<4$. Since the leading
density goes with $n^{10/3}$ it will
dominate over the positive kinetic part which goes like
$n^{2/3}$. We find that for densities around $3 n_0$ the total
energy has a maximum and starts to decrease continuously for higher
densities. The maximal
energy is changed towards higher values if we now include three
- particle correlations since the latter remain positive and have the same
leading density term as the two - particle correlational energy.

The three - particle part  reads
\be
&&E_{\rm corr_3}(\infty)=-g
     \int \frac{dp_1 dp_2 dp_3 dp_1' dp_2'dp_3'}{(2 \pi \hbar
)^{15}}
     T_3^2
          \nonumber \\
  &\times& \delta (p_1 + p_2 +p_3- p_1' - p_2'-p_3')
f_1' f_2' f_3'
{\bar f_1}
{\bar f_2}
{\bar f_3}
     \nonumber \\
  &\times & {{\cal P} \over E_1 + E_2 +E_3- E_1' - E_2'-E_3'}.
\label{eq3}
\ee
The analytic result for the contact potential
has not been given in the literature to our knowledge and reads
\be
{E_{\rm corr_3}\over n}&=&4  \epsilon_f {9013\over 2 \cdot 9 \cdot 25
  \cdot 77 \cdot 13} ({p_f^4 m T_3\over 4 \pi^4 \hbar^6})^2\nonumber\\
&=&{2.37 10^{-6}\over {\rm MeV\, fm}^2} n^{10/3} t_3^2.
\label{e3}
\ee
This three - particle correlation energy remains positive and has the
same leading density behaviour of $n^{10/3}$ as the two - particle
correlational energy. Since the pre-factor is smaller than the one for
the two - particle correlational energy we obtain a maximum at
3 times nuclear density beyond which the two - particle part dominates and
the total energy diverges negatively which would mean a collapse of
the system.
This clearly marks the limit of the Born approximation.

\begin{figure}
\psfig{file=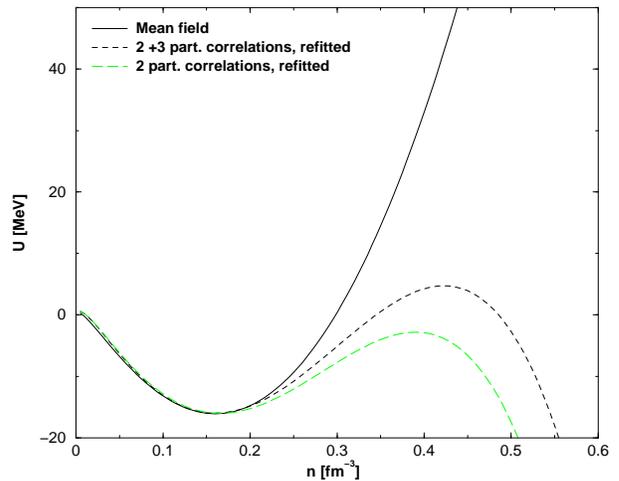,width=8cm,angle=-90}
\caption{The equation of state due to mean field compared with the
  equation of state including two - and three - particle collisions in
  Born approximation. }\label{1}
\end{figure}

Taking now the correlational energy into account via (\ref{hf3}) instead of only
the Hartree Fock energy 
(\ref{hf2}) we obtain a fit to the nuclear binding
of
\be
\tilde t_0&=&    -745.71{\rm MeV \, fm^3}\nonumber\\
\tilde t_3&=&8272.8.{\rm MeV \, fm^6}
\label{fit1}
\ee
which leads to a compressibility of
\be
\tilde K=351 {\rm MeV}
\ee
which is somewhat lower than the mean field compressibility of (\ref{k}).
The comparison of the two equation of states with and without two - and
three-particle correlations can be seen in figure \ref{1}. The inclusion of two - particle correlation energy
leads to a maximum at 3 times nuclear density above which the system
collapses. The complete result including three particle correlational
energy leads to a higher reachable maximum.

\subsection{Importance of three particle collisions}

We have seen that the three - particle contact interaction induces in a
natural way a density dependence of the two - particle cross section. 
Additionally we have obtained an explicit three - particle collision
integral on the same microscopic footing. We want now to answer the
question how important the three - particle collisions are. Therefore we
use
as a measure the ratio of the two- and three - particle ground state
correlation energies, (\ref{e2}) and (\ref{e3}), since this gives the
measure of how much energy is maximally dispersed by the corresponding
integrals.  We obtain
\be
\left |{E_2\over E_3}\right |&=&{2^5\cdot5\cdot 11\cdot 13 \over 9013}
(11-2\log{2})\nonumber\\
&\times&\left [9 {t_0^2 \over n^2 t_3^2}+\frac 3 4 + 5 {t_0 \over t_3 n} \right ].
\ee

This ratio is decreasing until it reaches the density 
\be
n_{\rm min}=-{18 t_0\over 5 t_3} =0.32 {\rm fm}^{-3}
\ee
where the ratio has the minimum
\be
\left |{E_2\over E_3}\right |_{\rm min}=1.4.
\ee
For higher densities the constant value
\be
\left |{E_2\over E_3}\right |_{\infty}=18.3
\ee
is approached.

I other words this means that the importance of three particle
collisions
increase with increasing density up to twice the nuclear density where
the correlational energies are nearly equal. For higher densities we
have 18 times larger two -particle correlational energies than three
-particle ones. At nuclear saturation density this ratio is
\be
\left |{E_2\over E_3}\right |_{n_0}=19.2
\ee
indicating that the three -particle collisions become important
between nuclear density and twice the nuclear density while it can be
neglected in the other cases.

\section{Conclusions}
 
For a microscopic two - and three - particle contact interaction
consisting of two parameters the
Kadanoff and Baym equation of motion is derived. From this a Boltzmann
kinetic equation is obtained with drift which turns out to be the known Skyrme
Hartree Fock expression while the collision side consists of two -
and three - particle collision integrals. The two - particle collision
integral contains an explicit density - dependent cross section arising
from the three - particle contact interaction. By this way both the
drift side as well as collision side is derived from the same
microscopic footing and no hybrid assumptions about separate density
dependent mean field and cross section is needed.

The correlational energy
for the three- and the two - particle part are calculated analytically.
Due to the density dependent two - particle collision integral both
correlational energies have the same leading power of
$(n/n_0)^{10/3}4$ in density which
shows that both contributions are of the same importance if three - particle
interaction has to be considered. This is clearly motivated by
the saturation point where non - relativistic two - particle approaches
fail to overcome the ``Coester line''. 

We find that there is a maximum in the energy at 3 times nuclear
density if two - and three
particle correlational energies are included. Beyond this density the
Born approximation fails at least in that the system collapses towards
diverging negative energy.

While the two - particle collision cross section has a natural density
dependence due to three - particle contact interaction it turns out that
the explicit three particle collision integral can be neglected as
long as one is below nuclear density. Around twice nuclear density the
three particle collision integral has the same importance as the two -      
particle one since it disperses the same amount of energy. For higher
densities the three - particle collision integral is again negligible.

I would like to thank H.S. K{\"o}hler and P. Lipavsky for numerous
discussions and the LPC for a friendly and hospitable
atmosphere. P. Chocian is thanked for reading the manuscript.

%\bibliography{kmsr,kmsr1,kmsr2,kmsr3,kmsr4,kmsr5,kmsr6,kmsr7,delay2,spin,refer,delay3,gdr}

\begin{thebibliography}{10}

\bibitem{K74}
H.~S. K{\"o}hler, Phys. Rep. {\bf 18},  217  (1975).

\bibitem{HMPO96}
M. Hjorth-Jensen, H. Muether, A. Polls, and E. Osnes, J. Phys. G {\bf 22},  321
   (1996).

\bibitem{S56}
T.~H.~R. Skyrme, Phil. Mag. {\bf 1},  1043  (1956).

\bibitem{S59}
T.~H.~R. Skyrme, Nucl. Phys. {\bf 9},  615  (1959).

\bibitem{CPW83}
J. Carlson, V. Pandharipande, and R. Wiringa, Nucl. Phys. A {\bf 401},  59
  (1983).

\bibitem{JRK83}
A. Jackson, M. Rho, and E. Krotscheck, Nucl. Phys. A {\bf 407},  495  (1983).

\bibitem{BFF99}
T. Gross-Boelting, C. Fuchs, and A. Faessler, Nucl. Phys. A {\bf 648},  105
  (1999).

\bibitem{VB72}
D. Vautherin and D.~M. Brink, Phys. Rev. C {\bf 5},  626  (1972).

\bibitem{BGM94}
A. Bonasera, F. Gulminelli, and J. Molitoris, Phys. Rep. {\bf 243},  1  (1994).

\bibitem{Ko75}
H.~S. K{\"o}hler, Nucl. Phys. A {\bf 258},  301  (1976).

\bibitem{BGM87}
A. Bouyssy, J.~F. Mathiot, and N.~V. Giai, Phys. Rev. C {\bf 36},  380  (1987).

\bibitem{BWBS87}
G.~E. Brown, W. Weise, G. Baym, and J. Speth, Comm. Nucl. Part. Phys. {\bf 17},
   39  (1987).

\bibitem{SRS85}
S. Sinha, J. Ram, and Y. Singh, Physica A {\bf 133},  247  (1985).

\bibitem{KB62}
L.~P. Kadanoff and G. Baym, {\em Quantum Statistical Mechanics} (Benjamin, New
  York, 1962).

\bibitem{LSV86}
P. Lipavsk{\'y}, V. {\v S}pi{\v c}ka, and B. Velick{\'y}, Phys. Rev. B {\bf
  34},  6933  (1986).

\bibitem{D90}
P. Danielewicz, Ann. Phys. (NY) {\bf 197},  154  (1990).

\bibitem{AYG98}
S. Ayik, O. Yilmaz, A. Golkalp, and P. Schuck, Phys. Rev. C {\bf 58},  1594
  (1998).

\bibitem{M94}
K. Morawetz, Phys. Lett. A {\bf 199},  241  (1995).

\bibitem{MK97}
K. Morawetz and H. K{\"o}hler, Eur. Phys. J. A {\bf 4},  291  (1999).

\bibitem{SLM96}
V. {\v S}pi{\v c}ka, P. Lipavsk{\'y}, and K. Morawetz, Phys. Lett. A {\bf 240},
   160  (1998).

\bibitem{LSM97}
P. Lipavsk{\'y}, K. Morawetz, and V. {\v S}pi{\v c}ka,   (1999), book sub. to
  Annales de Physique, K. Morawetz, Habilitation University Rostock 1998.

\bibitem{PN66}
D. Pines and P. Nozieres, {\em The Theory of Quantum Liquids} (Benjamin, New
  York, 1966), Vol.~1.

\bibitem{LL79}
E. Lifschitz and L.~P. Pitaevsky,  in {\em Physical Kinetics}, edited by E.
  Lifschitz (Akademie Verlag, Berlin, 1981).

\end{thebibliography}
%\bibliographystyle{prsty}

\end{document}